\documentclass[twocolumn,10pt]{revtex4}
\usepackage{bm,graphicx}
\usepackage[english]{babel}
\usepackage{amssymb, amsthm, amsmath}
\def\be{\begin{eqnarray}}

\def\ee{\end{eqnarray}}

\def\bee{\begin{eqnarray*}}

\def\eee{\end{eqnarray*}}

\newtheorem{thm}{Theorem}

\newtheorem{lem}{Lemma}

\newcommand{\finpr}{\hfill $\square$ \vspace{2mm}}

\begin{document}

\title{Ising models and topological codes: classical algorithms and quantum simulation}

\author{M. Van den Nest$^1$ and W.\ D\"ur$^2$}

\affiliation{ $^1$Max-Planck-Institut f\"ur Quantenoptik, Hans-Kopfermann-Str.~1, D-85748 Garching, Germany\\ $^2$ Institut f\"ur Theoretische Physik, Universit\"at
  Innsbruck, Technikerstr. 25, A-6020 Innsbruck,  Austria.}
\date{\today}

\begin{abstract}
We present an algorithm to approximate partition functions of 3-body
classical Ising models on two-dimensional lattices of arbitrary genus, in
the real-temperature regime. Even though our algorithm is purely classical,
it is designed by exploiting a connection to topological quantum systems,
namely the color codes. The algorithm performance is exponentially better
than other approaches which employ mappings between partition functions and
quantum state overlaps. In addition, our approach gives rise to a protocol
for quantum simulation of such Ising models by simply measuring local
observables on color codes.
\end{abstract}

%75.10.Hk    Classical spin models
%75.10.Pq    Spin chain models
%02.70.-c    Computational techniques
%03.67.Lx    Quantum computation

\maketitle

\textbf{Introduction.---}
In recent years, several new cross connections between classical spin models on lattices
and quantum information science have been discovered  \cite{Li97, Br07, Bo08, Va07, Va08, Yu10, Q_algorithms, Q_algorithms2, Ar10}. This line of research has lead to a transfer of knowledge and techniques between these two fields, yielding e.g. insights into measurement-based quantum computation \cite{Ra01,Br07, Bo08}, completeness results for classical models \cite{Va08}, but also new classical and quantum algorithms \cite{Q_algorithms, Q_algorithms2, Ar10, Va11}.

A particularly interesting connection has been established between topological quantum systems and the classical Ising model \cite{Br07, Va07, Bo08}. Topological quantum states, most prominently Kitaev's toric code \cite{Ki03}, have received tremendous attention recently. They constitute new phases of matter \cite{We04} and are essential for topological quantum computation \cite{Na08}. The connection between topological quantum systems and the classical Ising model is obtained by considering the Ising model partition function: the latter can be represented as the inner product between a topological quantum state and a certain product state \cite{Br07, Va07, Bo08}. Here the topological state encodes the geometry of the Ising model and the product state encodes interaction strengths and temperature.

In this paper we show how a classical-quantum correspondence can be exploited to design a new algorithm for estimating the partition function of Ising models. More precisely we consider Ising models with 3-body inhomogeneous interactions defined on two-dimensional lattices embedded in surfaces of arbitrary genus. Such models are known to be linked to the \emph{topological color codes} (TCCs) \cite{Bo08}. Even though our partition function algorithm is purely classical, it is obtained by utilizing the associated quantum formulation. The particular connection to TCCs exploited in the algorithm is a modification of the overlap mapping described above. Interestingly, by means of this modification, our algorithm offers an {\em exponential speedup} as compared to algorithms (both quantum and classical) that are based on using state overlaps directly \cite{Bo08, foot, Va11} (see also \cite{Ar10, foot2}). Furthermore, the algorithm applies to Ising models in a real-temperature regime. This is in contrast to other approaches that only work for complex i.e. unphysical temperatures \cite{Q_algorithms2}.

What is more, our approach also leads to an efficient \emph{quantum simulation} of such 3-body Ising models: the partition function can be estimated by measuring expectation values of certain simple local observables of the topological color code state. The different parameter regimes of the Ising model can be accessed by varying the measured local observables, providing an easy tool to study phase transitions or other interesting features of  3-body Ising models. Interestingly, TCC states can be prepared efficiently on a quantum computer (i.e. with a polynomial number of gates; however, also any other method of preparing such topologically protected systems immediately allows for an efficient quantum simulation of the corresponding classical spin model).

We finally remark that both our classical algorithm and quantum simulation are obtained by exploiting a particular symmetry of the TCC i.e. its \emph{self-duality}. As such, our method is not restricted to color codes, but applies to all classical spin models associated with self-dual quantum codes. Interestingly,  the toric code however does not fall into this class.

\begin{figure}[ht]
\begin{picture}(210,90)
\put(-35,-160)
{\includegraphics[width=12cm]{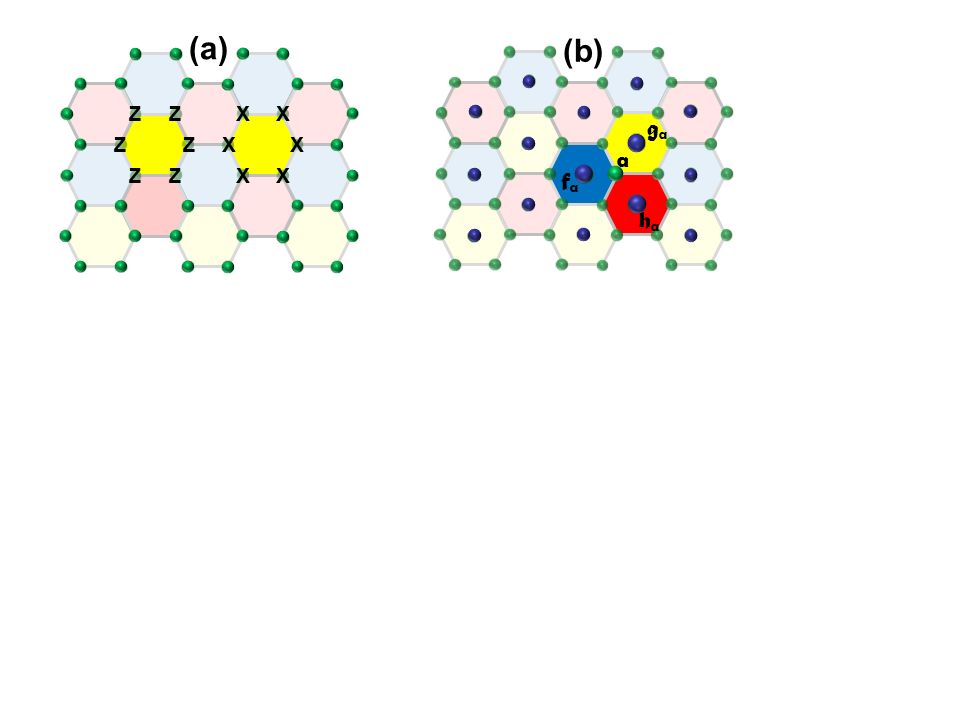}}
\end{picture}
\caption[]{\label{Fig_ColorCode} (a) Color code on a hexagonal lattice:
qubits are placed on vertices, and operators $X_{\mathfrak{f}}$ and $Z_{\mathfrak{f}}$ act on all qubits of a face. (b) Corresponding Classical Ising model: spins are placed on faces, and 3-body interactions take place between spins on faces $\mathfrak{f}_a$, $\mathfrak{g}_a$ and $\mathfrak{h}_a$  adjacent to vertex $a$.} \end{figure}

\textbf{Color codes.---}
Here we define TCCs \cite{Bo06}.  A 2-colex ${\cal C} $ is a 2-dimensional lattice embedded in a torus of arbitrary genus $g$, with the following properties: (a) every vertex of the lattice has degree 3, and (b) the faces of the lattice are 3-colorable. Examples  are the hexagonal lattice (Fig. \ref{Fig_ColorCode}) and the 4-8 lattice.

With every 2-colex we associate a Hilbert space by placing a qubit on each vertex. Denoting by $X$ and $Z$ the standard Pauli matrices, for every face $\mathfrak{f}$ we  define the following commuting operators (Fig \ref{Fig_ColorCode}): \be\label{TCC_text}
X_{\mathfrak{f}} :=\bigotimes_{v\in\mathfrak{f}} X_v \mbox{ and } Z_{\mathfrak{f}} :=\bigotimes_{v\in\mathfrak{f}} Z_v.
\ee
The associated TCC is  the space of all states $|\psi\rangle$ satisfying $X_{\mathfrak{f}}|\psi\rangle =|\psi\rangle = Z_{\mathfrak{f}}|\psi\rangle$ for all faces $\mathfrak{f}$.  Equivalently, this is the ground space of the Hamiltonian
\be\label{hamiltonian_text}
{\cal H}_{\mbox{\scriptsize{\sc tcc}}} = -\sum_{\mathfrak{f}} \ (X_{\mathfrak{f}} + Z_{\mathfrak{f}}).
\ee
It will be important in this work that this Hamiltonian is \emph{self-dual} in the sense that ${\cal H}_{\mbox{\scriptsize{\sc tcc}}}$ is invariant under the transformation $A\to HAH$ applied to each qubit separately, where $H$ denotes the Hadamard operation
(this  follows from the property $HXH=Z$). In general, the ground state space is degenerate. In this work we are interested in one particular ground state, namely \be\label{omega_text1} |\Omega\rangle =  N^{-1} \prod_{\mathfrak{f}} (I + X_{\mathfrak{f}})|0\rangle\ee where $N\geq 0$ is a normalization factor.

\textbf{3-Body Ising models.---}
Consider an arbitrary $2$-colex with vertex set ${\cal V}$ and face set ${\cal F}$. Place a classical Ising spin $\sigma_{\mathfrak{f}}\in\{1, -1\}$ on every face $\mathfrak{f}\in {\cal F}$. At each vertex, we consider a 3-body Ising interaction between the 3 spins located on the faces adjacent to each vertex (Fig. \ref{Fig_ColorCode}). More precisely, for every vertex $a\in {\cal V}$ let $\mathfrak{f}_a$, $\mathfrak{g}_a$ and $\mathfrak{h}_a$ denote the 3 faces adjacent to $a$ (labeled in no particular order). Then the  spins $\sigma_{\mathfrak{f}_a}$, $ \sigma_{\mathfrak{g}_a}$ and $ \sigma_{\mathfrak{h}_a}$ interact as $-J_a \sigma_{\mathfrak{f}_a}\sigma_{\mathfrak{g}_a}\sigma_{\mathfrak{h}_a}$ where $J_a$ is a (positive or negative) site-dependent interaction strength. Altogether, the energy of a spin configuration $\mathbf{\sigma}=\{\sigma_{\mathfrak{f}}: \mathfrak{f}\in {\cal F}\}$ is \be\label{H_ising_text} H(\mathbf{\sigma}) = -\sum_{a\in {\cal V}} J_a \sigma_{\mathfrak{f}_a}\sigma_{\mathfrak{g}_a}\sigma_{\mathfrak{h}_a}.\ee
The partition function of this model is ${\cal Z}= \sum_{\sigma} e^{-\beta H(\sigma)}$, and all relevant quantities of the classical model can be derived from ${\cal Z}$.

The connection between TCC states and the above Ising model is obtained as follows. Define
\be\label{alpha_gamma}
|\alpha_a\rangle &:=& \frac{e^{\beta J_a}|0\rangle_a + e^{-\beta J_a}|1\rangle_a}{\sqrt{e^{2\beta J_a} + e^{-2\beta J_a}}}\quad \mbox{with }a\in {\cal V}\nonumber \\
\gamma &:=& \sqrt{2^{F+2}} \ \prod_{a\in {\cal V}} \sqrt{e^{2\beta J_a} + e^{-2\beta J_a}}
\ee With the product state $|\alpha\rangle := \bigotimes_a |\alpha_a\rangle$ we then have
\be\label{overlap_text}
{\cal Z}= \gamma \cdot \langle \Omega|\alpha\rangle.
\ee
Thus the partition function is related (up to the easily computable prefactor $\gamma$) to the overlap between $|\Omega\rangle$ and the product state $|\alpha\rangle$, which contains information about temperature and couplings. The mapping (\ref{overlap_text}) was first proved in \cite{Bo08} in a slightly different form, i.e. using a different product state instead of $|\alpha\rangle$,  in which case it applies to surfaces of trivial topology ($g=0$). The identity (\ref{overlap_text}) however holds for arbitrary genus \cite{Sup} (Sec. \ref{sect_overlap}).

The overlap mapping (\ref{overlap_text}) can be used to compute the partition function ${\cal Z}$ by calculating the overlap $\langle \Omega|\alpha\rangle$. In fact, one can (approximately) compute such overlaps either on a quantum computer or on a classical computer. A natural quantum algorithm approach is to compute $\langle \Omega|\alpha\rangle$ via the Hadamard test. Since TCC states are stabilizer states on $V$ qubits, there exists a quantum circuit ${\cal C}$ of size poly$(V)$ such that $|\Omega\rangle = {\cal C}|0\rangle$ \cite{Go97}. Since $|\alpha\rangle$ is a product state, there also exists a poly-sized quantum circuit ${\cal C}'$ such that $|\alpha\rangle= {\cal C}'|0\rangle$. This leads to $\langle \Omega|\alpha\rangle= \langle 0|{\cal C}^{\dagger}{\cal C}|0\rangle$. Since the overall circuit ${\cal C}^{\dagger}{\cal C}'$ has polynomial size, the matrix element $\langle 0|{\cal C}^{\dagger}{\cal C}|0\rangle$ can be estimated on a quantum computer using the Hadamard test. This allows to estimate  $\langle \Omega|\alpha\rangle$ in polynomial time with error $1/$poly$(V)$ (with a success probability which is exponentially close to $1$).  Using ${\cal Z} = \gamma \cdot \langle \Omega|\alpha\rangle$, this leads to an estimate of ${\cal Z}$ with error
\be\label{deltaolap}
\Delta_{\mbox{\scriptsize old}}=\frac{\gamma}{\mbox{poly}(V)}.
\ee
An equivalent approach was taken in \cite{Ar10} for Ising models associated with the toric code state.

The quantity $\langle \Omega|\alpha\rangle$ may however also be estimated directly on a classical computer. This can be done using the probabilistic techniques from \cite{Va11}, where a general method was given to estimate overlaps between stabilizer states and product states. It turns out that the resulting classical algorithm allows to estimate  ${\cal Z}$ in polynomial time with the same error scaling  (\ref{deltaolap}) as achieved by the Hadamard test.

In the present paper we will provide a new classical algorithm to estimate ${\cal Z}$ which outperforms the above approach by an exponential factor, as shown next.

\textbf{Main results.---}
Let $V$ denote the number of vertices of the colex and $F$ the number of faces. We provide an algorithm with runtime \be\label{main_runtime_text} O( V^3\cdot \frac{1}{\epsilon^2} \cdot \log\frac{1}{1-p})\ee which outputs an estimate ${\cal Z}_{\mbox{\scriptsize{\sc{est}}}}$ of the partition function ${\cal Z}$, such that the inequality \be\label{estimate_quantum_text} |{\cal Z}_{\mbox{\scriptsize{\sc{est}}}}-{\cal Z}|\leq  \frac{\gamma}{ \sqrt{2^{F-2}}} \cdot \epsilon\ee is satisfied with probability at least $p$.
The algorithm applies to all temperature regimes and arbitrary inhomogeneous couplings $J_a$.

We now discuss the performance of our algorithm and compare it to other approaches. First, the 3-body Ising models considered in this work are exactly solvable (using the Bethe Ansatz) for hexagonal and 4-8 lattices on trivial topologies in the case of uniform couplings \cite{Ising_solvable}. Our result, however, is considerably more general in that it applies to arbitrary lattices on surfaces of arbitrary genus and for site-dependent couplings.

Second, the runtime scaling (\ref{main_runtime_text}) implies that in poly$(V)$ time it is possible to achieve  $\epsilon = 1/$poly$(V)$ and a success probability $p$ which is exponentially (in $V$) close to 1.  The error in estimating ${\cal Z}$ as given in (\ref{estimate_quantum_text}) thus scales as
\be\label{delta_new}
\Delta_{\mbox{\scriptsize}}=  \frac{\gamma}{\sqrt{2^{F-2}}\cdot \mbox{poly}(V)}.
\ee
We can now compare the performance of our algorithm with approaches that estimate state overlaps directly, giving rise to the approximation scaling (\ref{deltaolap}). Interestingly, the accuracy of our algorithm constitutes an \emph{improvement by a factor of $2^{-F/2+1}$} compared to the overlap approach. Note that this factor is \emph{exponentially small} in the number of sites $V$ of the lattice in many cases of interest. Indeed, it can be shown \cite{Sup} (Sec. \ref{sect_F}) that
\be\label{F_text}
F = \frac{V-4g}{2} + 2
\ee
where $g$ is the genus of the surface. Thus whenever the genus is not too large (e.g. $g$ constant, $g=O(\log V)$ or even any $g \leq  V/5$) we will have $F=O(V)$, in which case  $2^{-F/2+1}$ will be exponentially small and the error  $\Delta_{\mbox{\scriptsize}}$ is thus exponentially smaller than $\Delta_{\mbox{\scriptsize old}}$.

What is more, we find \cite{Sup} (Sec. \ref{sect_F}) that \be\label{bound_Z_text} {\cal Z}\leq \frac{\gamma}{\sqrt{2^{F-2}}}.\ee Hence, whenever $2^{-F/2+1}$  is exponentially small,  the approximation error $\Delta_{\mbox{\scriptsize old}}$ is exponentially larger than ${\cal Z}$ itself and thus meaningless. This demonstrates that, for such lattices, the overlap approach is  not useful for \emph{any} values of temperature and couplings. In short, we have demonstrated that our algorithm yields drastic improvements over the overlap approach.

An important question is to understand for which regimes of temperature and couplings the partition function ${\cal Z}$ is comparable in size to $\Delta_{\mbox{\scriptsize}}$. In particular, if $\Delta_{\mbox{\scriptsize}}$ turns out to be larger than ${\cal Z}$, the approximation provided by the algorithm is not useful. Note that it is hard to determine whether or not  such an issue occurs, since computing ${\cal Z}$ is hard---indeed this is the goal of the algorithm in the first place. This type of problem is not so much a drawback of our algorithm in particular, but is typical for all algorithms which, as ours,  provide so-called additive approximations (see \cite{Q_algorithms2}).
Here we show that, for certain instances, $\Delta_{\mbox{\scriptsize}}$ is indeed provably significantly smaller than ${\cal Z}$.
An example of such a case is $T=0$ and $J_{a}\geq 0$. This is the zero temperature regime in a ferromagnetic system. In this case one has  ${\cal Z}= \gamma/2^{F/2-1}$ \cite{Sup} (Sec. \ref{sect_high_low_T}) which is indeed much larger than $\Delta_{\mbox{\scriptsize}}$.  A second example is $T=\infty$.  In this case it can be shown \cite{Sup} (Sec. \ref{sect_high_low_T}) that  ${\cal Z}= \gamma2^{-F/2+1} 4^{-g}$. This is much larger than $\Delta_{\mbox{\scriptsize}}$ as long as the genus $g$ scales as $O(\log V)$. In addition to the above examples, further instances of meaningful approximations can be identified. In particular,  configurations of the parameters $\beta$ and $\{J_a\}$ which are sufficiently close to the aforementioned examples will also give rise to a meaningful approximation as well, using a continuity argument.

\textbf{Proof ingredients.---} The proof of our main result is given in \cite{Sup} (Secs. \ref{sect_proof_main_step1}, \ref{sect_proof_main_step2}). Here we outline the main steps. {\it Step 1.} First we will prove a new identity relating the overlap $\langle\Omega|\alpha\rangle$ to a certain quantum expectation value. More precisely, using the shorthand notation
$|\alpha_a\rangle = x_a|0\rangle + y_a|1\rangle$ where $|\alpha_a\rangle$ is as in (\ref{alpha_gamma}), we introduce the tensor product operator
\be
A:= \bigotimes A_a, \quad \mbox{where} \quad A_a := \left[ \begin{array}{cc}x_a & y_a \\ y_a & -x_a\end{array} \right].
\ee
We will show that
\be\label{overlap_expectation_text}
\sqrt{2^{F - 2}} \cdot \langle \Omega| \alpha\rangle = \langle \Omega| A|\Omega\rangle.
\ee
This  identity is proved by exploiting the \emph{self-duality} of color codes. This self-duality implies that the state $|\Omega\rangle$ is an equal superposition state  of the form\be\label{omega_text} |\Omega\rangle \propto \sum_{s\in S} |s\rangle\ee where, crucially,  the set $S\subseteq\mathbb{Z}_2^V$ is a classical \emph{self-orthogonal linear code}. It is the self-orthogonality of $S$ which will allow to relate the overlap $\langle \Omega| \alpha\rangle$ to the expectation value $\langle \Omega| A|\Omega\rangle$. Interestingly, this seems to be a specific feature of color codes.

Combining (\ref{overlap_expectation_text}) and (\ref{overlap_text}) yields a relation between the partition function and the expectation value $\langle \Omega| A|\Omega\rangle$:\be\label{partition_expectation_text}  {\cal Z}=  \frac{\gamma}{ \sqrt{2^{F - 2}}}\cdot \langle \Omega| A|\Omega\rangle.\ee
Note that this identity   has a different character from the overlap mapping (\ref{overlap_text}). Indeed,   writing out the overlap $\langle\Omega|\alpha\rangle = \sum \langle \Omega|x\rangle\langle x|\alpha\rangle$ yields a direct term-by-term correspondence to the partition function ${\cal Z}$ in the sense that each term $\langle \Omega|x\rangle\langle x|\alpha\rangle$ is positive and immediately identified a Boltzmann weight of the corresponding Ising model (up to the multiplicative factor $\gamma$). In contrast, the expansion $\langle \Omega| A|\Omega\rangle = \sum  \langle\Omega|x\rangle\langle y|\Omega\rangle \langle x| A|y\rangle$ contains both positive an negative terms. Thus a nontrivial recombination of terms occurs to ensure that the resulting sum indeed yields a proper partition function.

{\it Step 2.} We will show that the quantum expectation value $\langle\Omega| A|\Omega\rangle$ can be approximated efficiently on a \emph{classical} computer with error $\epsilon$ with runtime (\ref{main_runtime_text}).
To achieve this, we will use methods for classically simulating quantum systems. A key ingredient is that each matrix $A_a$, which is a real orthogonal matrix with determinant $-1$,  can be decomposed as $A_a=ZU^{\dagger}D_aU$; here $Z$ is the standard Pauli matrix, $U= HP$ where $H$ is the Hadamard gate and $P=$ diag$(1, i)$, and  $D_a$ is a diagonal matrix.  As a result, we find
\be\label{expectation_classical_algo_text}
\langle \Omega| A|\Omega\rangle = \langle\Omega| Z^{\otimes V}[U^{\otimes V}]^{\dagger} \bigotimes D_a U^{\otimes V}|\Omega\rangle.
\ee
Furthermore $Z^{\otimes V}|\Omega\rangle = |\Omega\rangle$ so that the operator $Z^{\otimes V}$ can be absorbed. Writing  $|\varphi\rangle := U^{\otimes V} |\Omega\rangle$ and $f(x):= \langle x|\bigotimes D_a|x\rangle $ yields  \be \langle \Omega| A|\Omega\rangle = \sum  f(x)  |\langle x|\varphi\rangle|^2\equiv \langle f\rangle.\ee This is the expectation value of a random variable on the set of bit strings $x$ which takes the value $ f(x)$ with probability $|\langle x|\varphi\rangle|^2$. The expectation value $\langle f\rangle$ can hence be estimated by sampling this random value; the number of samples needed to get an error $\epsilon$ is $O(1/\epsilon^2)$. Apart from the number of samples, one needs to take into account the complexity of generating each sample as well as the complexity of computing $f(x)$.  The following property is now crucial: since  $U$ is a Clifford operation and the TCC states are stabilizer states, also $|\varphi\rangle$ is a stabilizer state. The Gottesman-Knill theorem then allows to sample the distribution $\{|\langle x|\varphi\rangle|^2\}$ efficiently, more precisely in $O(V^3)$ time \cite{Go99}. Furthermore, its easy to show that the values $f(x)$ can be computed efficiently (more precisely: in $O(V)$ time). Putting together the time complexity of the sampling, the time complexity of computing $f(x)$ and the number of samples  yields a classical algorithm with scaling (\ref{main_runtime_text}) to provide an $\epsilon$-approximation of $\langle\Omega|A|\Omega\rangle$. Using (\ref{partition_expectation_text}) then also yields a classical algorithm for estimating ${\cal Z}$ with error (\ref{estimate_quantum_text}).

\textbf{Quantum Simulation.---}
The identity (\ref{partition_expectation_text}) also gives rise to an immediate method for a quantum simulation algorithm of the classical 3-body Ising models, i.e. to estimate ${\cal Z}$. The simulation simply consists of preparing $|\Omega\rangle$ and measuring the expectation value of the local observable $A$ which encodes coupling strengths and temperature of the model. The state $|\Omega\rangle$ can be generated efficiently on a quantum computer. Since TCC states are stabilizer states, a poly-sized quantum circuit with $O(V^3)$ gates suffices \cite{Go97} (however, any other method to prepare color codes, e.g. as ground states of some effective Hamiltonian, is also suitable).
Furthermore $A$ can be measured in $O(V)$ time---in fact all individual observables $A_a$ can be measured simultaneously. Using standard probability theory bounds,  this results in a quantum simulation that allows one to estimate $\langle \Omega| A|\Omega\rangle$ with probability $p$ and error $\epsilon$ with runtime  (\ref{main_runtime_text}).  Using (\ref{partition_expectation_text}) then immediately yields a quantum simulation algorithm for estimating ${\cal Z}$ with runtime (\ref{main_runtime_text}). Thus the quantum simulation performance  is the same as that of the classical algorithm. Nevertheless, the quantum simulation might be appealing in its own right, e.g. owing to the simplicity of the protocol.

\textbf{Conclusion and outlook.---}
We have presented a classical algorithm for the simulation of 3-body Ising models associated with topological quantum systems. The algorithm was constructed via a detour, i.e. by classically simulating properties of quantum systems. It is capable of simulating inhomogeneous models on lattices with arbitrary genus, and can be applied to systems in the real-temperature regime. This opens the way to efficiently simulate such systems and investigate their properties.
At the same time, our approach gives rise to an efficient quantum simulation of the classical model that only involves measurement of local observables on a topological color code state. We have shown that the achieved approximation (both for the classical algorithm and quantum simulation) is meaningful in the low and high temperature limit. Furthermore our method gives rise to an exponential improvement as compared to previous approaches.  It is interesting to note that our techniques do not apply to 2-body Ising models, which are associated with  Kitaev's toric code  via an analogous overlap mapping. Despite the fact that the overlap mapping is very similar, the underlying classical code is in the case of the toric code not self-orthogonal which prevents the application of the techniques established in this paper, i.e. rewriting state overlaps as expectation values. The same fact also prevents a direct quantum simulation. Whether our methods can be generalized to other models remains an open problem.

Finally, we mention that the techniques presented in this work (in particular Eq. (\ref{overlap_expectation_text})) can be used in a rather different area, namely to compute the \emph{geometric measure of entanglement} in TCC states. This is done elsewhere \cite{Or13}.

\textit{Acknowledgements.---}
This work was supported by the Austrian Science Fund (FWF): P24273-N16, SFB F40-FoQus F4012-N16. We thank G. de las Cuevas, M. A. Martin-Delgado and H. Bombin for helpful discussions.

\newpage

\

\newpage

\begin{center}
{{\bf\Large\sc Supplementary Material}}
\end{center}
Here we provide mathematical proofs of our results. The structure of this Supplementary Material is as follows.  In section \ref{sect_preliminary} we introduce preliminary concepts.  In section \ref{sect_TCC_ising} we prove the overlap mapping Eq. (\ref{overlap_text}). In sections \ref{sect_proof_main_step1} and \ref{sect_proof_main_step2} we prove our main results (Step 1 and Step 2, respectively). In section \ref{sect_quantum_simulation} we provide our quantum simulation algorithm. In section \label{sect_F} we prove Eqs. (\ref{F_text}) and  Eq. (\ref{bound_Z_text}). Finally, in section \ref{sect_high_low_T} we discuss the high and low temperature behavior of the 3-body Ising partition function as discussed in the last paragraph of the \emph{Main Results} section.

\section{Preliminary concepts}\label{sect_preliminary}
In this section we introduce the basic concepts that will be central in the proofs of our results. After fixing some basic notation in section \ref{sect_notation} we introduce topological color codes (TCCs) in section \ref{sect_TCC}.
In section \ref{sect_CSS} we recall basic properties of Calderbank-Shor-Steane (CSS) codes, of which TCCs are examples.

\subsection{Notation}\label{sect_notation}
We will denote by $X$ and $Z$ the standard Pauli matrices. If $u=(u_1, \dots, u_n)\in\mathbb{Z}_2^n$ is an $n$-bit string, we will often consider the following $n$-qubit tensor product operators: \be\label{X_and_Z} X(u)&:=& X^{u_1}\otimes\dots\otimes X^{u_n}\nonumber \\ Z(u)&:=& Z^{u_1}\otimes\dots\otimes Z^{u_n}\ee where $X^1=X$ and $X^0=I$ and similar for $Z$. We call $X(u)$ an $X$-type operator and $Z(u)$ a $Z$-type operator, respectively. It is easily verified that, for every $u, v\in \mathbb{Z}_2^n$, we have \be X(u)|v\rangle &=& |v+u\rangle\nonumber\\ Z(u)|v\rangle &=& (-1)^{u^Tv}|v\rangle\nonumber\\ X(u)Z(v)&=& (-1)^{u^Tv}Z(v)X(u)\label{X_and_Z_properties}\ee where $|v\rangle$ denotes an $n$-qubit standard basis state in the usual sense and where $u+v$ is computed over $\mathbb{Z}_2$.

\subsection{Topological color codes}\label{sect_TCC}

Recall the definition of a 2-colex ${\cal C} $ given in the main text. Let ${\cal V}$ and ${\cal F}$ denote the sets of faces and vertices and $F$ and $V$ denote the number of faces and vertices, respectively. The following two properties  easily follow from the definition of 2-colexes:

\begin{lem}{\bf (Basic properties of 2-colexes)}\label{thm_basic_properties_colex}
Consider an arbitrary 2-colex. Then every face contains an even number of vertices. Furthermore, every two distinct faces overlap in precisely two vertices, or do not overlap at all.
\end{lem}

Consider the TCC associated with the colex ${\cal C}$. The operators $X_{\mathfrak{f}}$ and $Z_{\mathfrak{f}}$ (recall (\ref{TCC_text}) in the main text) are called ($X$-type and $Z$-type, resp.) face operators. It is known that there are precisely $F-2$ independent $X$-type face operators (and hence $F-2$ independent $Z$-type operators as well) \cite{Bo06}. The dimension of the TCC (number of encoded qubits) is $2^m$ where \be m &=& \# \mbox{ qubits} - \# \mbox{ independent stabilizers}\\
&=& V - 2(F-2)\\ &=& 4-2\chi = 4g,\label{dimension}\ee where $\chi= F+ V-E$ is the Euler characteristic of the surface and where in the last identity we have used that $\chi = 2-2g$.

In this work we will be  interested in the  state $|\Omega\rangle$ as given in (\ref{omega_text1}) in the main text. This state is one of the ground states of ${\cal H}_{\mbox{\scriptsize{\sc tcc}}}$ i.e. one has \be X_{\mathfrak{f}}|\Omega\rangle =|\Omega\rangle = Z_{\mathfrak{f}}|\Omega\rangle\ee for all faces $\mathfrak{f}$. In the following it will also be convenient to work with the unnormalized state $|\Omega'\rangle := N|\Omega\rangle$.

Next we derive an explicit expansion of $|\Omega\rangle$ in the computational basis. To do so, define the $V\times F$ vertex-face incidence matrix $B$, which describes the incidence relation between faces and vertices of the colex, as follows: the rows of $B$ are indexed by vertices, the columns are indexed by the faces; the matrix element $B_{v, \mathfrak{f}}$ is defined by \be B_{v, \mathfrak{f}} = \left\{\begin{array}{cl} 1& \mbox{ if } v\in \mathfrak{f}\\ 0 & \mbox{ otherwise.}\end{array} \right.\ee  Equivalently, each column of $B$ is a 0/1 vector where the 1s occur to those vertices which comprise a given face.
Let $\mathbb{Z}_2^{\cal F}$ denote the set of 0/1 vectors with entries labeled by the faces: $t=(t_{\mathfrak{f}}: \mathfrak{f}\in{\cal F})$ with $t_{\mathfrak{f}}\in \mathbb{Z}_2$. With these definitions and using (\ref{X_and_Z_properties}), we find\be \prod_{\mathfrak{f}} (I + X_{\mathfrak{f}}) = \sum_{t\in\mathbb{Z}_2^{\cal F}} \prod_{\mathfrak{f}} X_{\mathfrak{f}}^{t_{\mathfrak{f}}} \ee and \be \prod_{\mathfrak{f}} X_{\mathfrak{f}}^{t_{\mathfrak{f}}}|0\rangle =  |Bt\rangle\ee  and hence \be |\Omega'\rangle &=& \prod_{\mathfrak{f}} (I + X_{\mathfrak{f}})|0\rangle\nonumber\\  &=& \sum_{t} \prod_{\mathfrak{f}} X_{\mathfrak{f}}^{t_{\mathfrak{f}}}|0\rangle = \sum_t |Bt\rangle. \label{Omega'}\ee
Finally, the set \be\label{S} S:= \{Bt: t\in \mathbb{Z}_2^{\cal F}\}\ee is a linear subspace of $\mathbb{Z}_2^{\cal V}$: for every $s, s'\in S$  it follows that $s+s'\in S$ where the sum $s+s'$ is computed over $\mathbb{Z}_2$. Since there are $F-2$ independent face operators $X_{\mathfrak{f}}$, there are $F-2$ linearly independent columns in the matrix $B$. Hence the cardinality of $S$ is $|S|=2^{F-2}$. Putting everything together, we arrive at the following identities; \be\label{psi_psi'} N^2&=& \langle\Omega'|\Omega'\rangle = 2^{F+2}\\ |\Omega\rangle& =& \frac{1}{\sqrt{2^{F+2}}} \sum_t|Bt\rangle\\ &=&\frac{1}{\sqrt{|S|}} \sum_{s\in S}|s\rangle\label{CSS}\ee

\subsection{CSS codes and states}\label{sect_CSS}

A linear subspace $C\subseteq \mathbb{Z}_2^n$ is called a (classical) binary linear code of length $n$. The elements of $C$ are called its codewords.  The orthogonal complement of $C$ is the set \be C^{\perp}:=\{v\in \mathbb{Z}_2^n: u^Tv =0 \mbox{ for all } u\in C\}\ee  which is also a linear code. A Calderbank-Shor-Steane (CSS) quantum code is defined as follows \cite{Go97}. Let $C$ and $D$ be two binary linear codes of length $n$ such that $D\subseteq C^{\perp}$. The CSS quantum code associated to the pair $(C, D)$ is the space of all $n$-qubit states $|\psi\rangle$ satisfying \be X(u)Z(v)|\psi\rangle = |\psi\rangle \mbox{ for every } u\in C, v\in D.\ee The set of operators \be{\cal S}:= \{X(u)Z(v) : u\in C, v\in D\}\ee is a commuting group called the stabilizer of the code. A generating set of ${\cal S}$ is obtained as follows. Let \be \{u^1, \dots, u^k\}\subseteq C \mbox{ and } \{v^1,\dots, v^{l}\}\subseteq D\ee be bases of $C$ and $D$, respectively, and denote \be \sigma_i:= X(u^i)\quad \mbox{and}\quad  \tau_j:=Z(v^j).\ee Then the operators $\{\sigma_1, \dots, \sigma_k, \tau_1, \dots, \tau_l\}$ form a generating set of ${\cal S}$.

For every binary linear code $C$ of length $n$, define an $n$-qubit state \be|C\rangle:=\frac{1}{\sqrt{|C|}}\sum_{u\in C}|u\rangle.\ee Any state of this kind is called a CSS state.  One can show that, for every $u\in C$ and $v\in C^{\perp}$, we have \be\label{CSS_stabilizer} X(u)Z(v)|C\rangle = |C\rangle.\ee Furthermore $|C\rangle$ is the unique state (up to a global phase) satisfying the equations (\ref{CSS_stabilizer}). Thus a CSS state is a one-dimensional CSS code.

Finally, coming back to color codes, it follows from the discussion in section \ref{sect_TCC} that every TCC is a CSS code associated with the classical codes $C\equiv S\equiv D$, where $S$ was defined in (\ref{S}).
Identity (\ref{CSS}) shows that $|\Omega\rangle$ is a CSS state; more precisely, we have $|\Omega\rangle = |S\rangle$.

\section{Proof of the overlap mapping (\ref{overlap_text})}\label{sect_TCC_ising}\label{sect_overlap}

Recall the definition of the vertex-face incidence matrix $B$ and consider a vector $Bt$ where $t=(t_{\mathfrak{f}}: \mathfrak{f}\in{\cal F})$. The  vector $Bt$ has one entry per vertex of the colex. For a vertex $a$, let $\mathfrak{f}_a$, $\mathfrak{g}_a$ and $\mathfrak{h}_a$ denote the 3 faces adjacent to $a$ as described in the main text. Then the $a$-th entry of $Bt$ is given by  the sum \be t_{\mathfrak{f}_a} + t_{\mathfrak{g}_a} + t_{\mathfrak{h}_a}\ee Recalling the expression (\ref{Omega'}) for $|\Omega'\rangle$, it follows that \be |\Omega'\rangle = \sum_t \bigotimes_a | t_{\mathfrak{f}_a} + t_{\mathfrak{g}_a} + t_{\mathfrak{h}_a}\rangle.\ee
Now define $|\alpha_a'\rangle := e^{\beta J_a}|0\rangle_a + e^{-\beta J_a}|1\rangle_a$. We claim that \be\label{overlap_psi'} {\cal Z} = \langle \Omega'|\bigotimes_a|\alpha_a'\rangle.\ee where ${\cal Z}$ is the partition function of the 3-body Ising model defined on ${\cal C}$ as defined in the main text. To prove this overlap relation, consider an arbitrary $0/1$ vector $t\in\mathbb{Z}_2^{\cal F}$ with entries indexed by the faces of the colex. Such a  $t$ corresponds directly to a configuration of the Ising spins; in particular, define $\sigma_{\mathfrak{f}}:= (-1)^{t_{\mathfrak{f}}} $ i.e. $t_{\mathfrak{f}} = 0$ iff $\sigma_{\mathfrak{f}}=1$. With these notations, we have \be \langle t_{\mathfrak{f}_a} + t_{\mathfrak{g}_a} + t_{\mathfrak{h}_a}| \alpha_a'\rangle &=& \left\{ \begin{array}{cl} e^{\beta J_a} & \mbox{ if } t_{\mathfrak{f}_a} + t_{\mathfrak{g}_a} + t_{\mathfrak{h}_a}=0\\ e^{-\beta J_a} & \mbox{ if } t_{\mathfrak{f}_a} + t_{\mathfrak{g}_a} + t_{\mathfrak{h}_a}=1 \end{array}\right.\nonumber\\ \nonumber \\&=&
\exp[\beta J_a \sigma_{\mathfrak{f}_a}\sigma_{\mathfrak{g}_a}\sigma_{\mathfrak{h}_a}] \ee
This implies that
\be \prod_a \langle t_{\mathfrak{f}_a} + t_{\mathfrak{g}_a} + t_{\mathfrak{h}_a}| \alpha_a'\rangle = \prod_a \exp[\beta J_a  \sigma_{\mathfrak{f}_a}\sigma_{\mathfrak{g}_a}\sigma_{\mathfrak{h}_a}].\ee The identity (\ref{overlap_psi'}) follows.

Finally, we rewrite the identity (\ref{overlap_psi'}) in terms of properly normalized states. For every vertex $a$ define \be \label{alpha_a} |\alpha_a\rangle:= \frac{|\alpha_a'\rangle}{\| |\alpha_a'\rangle\|} =  \frac{|\alpha_a'\rangle}{ \sqrt{e^{2\beta J_a} + e^{-2\beta J_a}}}\ee and denote $|\alpha\rangle= \bigotimes |\alpha_a\rangle$. Recall the identity (\ref{psi_psi'}) relating the unnormalized state $|\Omega'\rangle$ with the properly normalized state $|\Omega\rangle$. Together with (\ref{overlap_psi'}) we find:

\begin{thm}{\bf (Overlap mapping)}\label{thm_overlap}
Consider an arbitrary TCC on a 2-colex. Then \be\label{overlap2} {\cal Z}= \gamma \cdot \langle \Omega|\alpha\rangle \ee where \be \gamma = \sqrt{2^{F+2}} \ \prod_{a\in {\cal V}} \sqrt{e^{2\beta J_a} + e^{-2\beta J_a}}\ee
\end{thm}

Theorem \ref{thm_overlap} is related to a similar result obtained in \cite{Bo08}. There the authors obtain a relation of the form ${\cal Z}= \delta \cdot \langle \Omega|\tilde\alpha\rangle$ where $\delta$ is an easy-to-compute prefactor as $\gamma$ in theorem \ref{thm_overlap} and where $|\tilde\alpha\rangle = \bigotimes |\tilde\alpha_a\rangle$ is a product state where \be |\tilde\alpha_a\rangle = \cosh \beta J_a|0\rangle + \sinh \beta J_a|1\rangle.\ee  Note the difference between the product states $|\alpha\rangle$ and $|\tilde\alpha\rangle$. In contrast to theorem \ref{thm_overlap}, which holds for arbitrary 2-colexes, the relation ${\cal Z}= \delta \cdot \langle \Omega|\tilde\alpha\rangle$ obtained in \cite{Bo08} only holds for those colexes for which the state $|\Omega\rangle$ is the unique ground state. This means that the dimension of the code is 1, corresponding to $g=0$ or, equivalently,  $\chi=2$ owing to (\ref{dimension}). The relation between Ref. \cite{Bo08} and theorem \ref{thm_overlap} is obtained as follows. If the TCC has $|\Omega\rangle$ as its unique ground state, the self-duality of the TCC Hamiltonian ${\cal H}_{\mbox{\scriptsize{\sc tcc}}}$ implies that the state $|\Omega\rangle$ must be self-dual as well, i.e. $H^{\otimes V}|\Omega\rangle = |\Omega\rangle$. Noting further that \be H|\alpha_a\rangle \propto |\tilde\alpha_a\rangle,\ee the identity (\ref{overlap2}) implies that \be {\cal Z}= \gamma \cdot \langle \Omega|\alpha\rangle = \langle \Omega|H^{\otimes V}|\alpha\rangle = \delta\cdot \langle \Omega|\tilde\alpha\rangle \ee for some $\delta$.

\section{Proof of main result: step 1}\label{sect_proof_main_step1}

Next we prove identity (\ref{partition_expectation_text}) in the main text. To do so, in section \ref{sect_self_orthogonal} we first show that TCC states are CSS states for which the underlying classical codes are self-orthogonal. This property will then be used to prove (\ref{partition_expectation_text}) in section \ref{sect_expectation_value}.

\subsection{TCCs and self-orthogonal classical codes}\label{sect_self_orthogonal}

Here we show that the TCC state $|\Omega\rangle = |S\rangle$ is a CSS state of a special kind: the code $S$ is \emph{self-orthogonal}. This means that every two codewords are orthogonal: $s^Tt=0$ for every $s, t\in S$. Equivalently, $S\subseteq S^{\perp}$.

\begin{lem}{\bf (Self-orthogonality)}\label{thm_self_orthogonal}
The code (\ref{S})  is self-orthogonal.
\end{lem}
{\it Proof:} to show that $S$ is self-orthogonal it suffices to show that $B^TB=0$, that is, every two columns of $B$ are orthogonal and every column is orthogonal to itself over $\mathbb{Z}_2$. But this immediately follows from lemma \ref{thm_basic_properties_colex}: first, since every face has an even number of vertices, it follows that each column of $B$ is orthogonal to itself; second, since every two distinct faces either overlap in two sites or do not overlap at all, it follows that every two distinct columns of $B$ are orthogonal. \finpr

The self-orthogonality of $S$ is closely related to the self-duality of the color code Hamiltonian $H_{\mbox{\scriptsize{\sc tcc}}}$. Indeed we have the following general result. Consider a CSS code with stabilizer generators $\sigma_1, \dots, \sigma_k, \tau_1, \dots, \tau_l$ where each $\sigma_i$ is an $X$-type operator and each $\tau_j$ is a $Z$-type operator. Define the code Hamiltonian \be {\cal H}_{\mbox{\scriptsize{\sc code}}} = -\sum \sigma_i - \sum \tau_j\ee and define the state \be |\psi\rangle =  N^{-1} \prod_{i=1}^k (I + \sigma_i)|0\rangle\ee in analogy with the TCC state $|\Omega\rangle$. Then the following holds:

\begin{lem}{\bf(Self-duality and self-orthogonality)}
If  ${\cal H}_{\mbox{\scriptsize{\sc code}}}$  is self-dual then $|\psi\rangle$ is a CSS state where the associated classical code is self-orthogonal.
\end{lem}
{\it Proof:} write $\sigma_i= X(u^i)$ for some $u^i\in \mathbb{Z}_2^n$. Since the code Hamiltonian is self-dual, it follows that $k=l$ and (after possible relabeling of the stabilizers) $\tau_i= Z(u^i)$. Let $C\subseteq\mathbb{Z}_2^n$ be the code generated by the $u^i$. Using an argument analogous to the derivation of (\ref{CSS}) one shows that $|\psi\rangle = |C\rangle$. Since $\tau_i|\psi\rangle = |\psi\rangle$, it follows that $Z(u^i)|C\rangle = |C\rangle$.  This last identity implies (using (\ref{X_and_Z_properties})) that \be \sum_{v\in C} (-1)^{v^Tu^i}|v\rangle = \sum_{v\in C}|v\rangle.\ee It follows that $v^Tu^i=0$ for all $v\in C$ and for all $i$. Since the $u^i$ generate the code $C$, it follows that $v^Tu=0$ for all $u, v\in C$. Hence $C$ is self-orthogonal. \finpr

Finally, we will need the following basic property.
\begin{lem}\label{thm_Z_otimes_n}
Let $C$ be a self-orthogonal binary linear code of length $n$. Then $Z^{\otimes n}|C\rangle= |C\rangle$.
\end{lem}
{\it Proof: } let $d\in\mathbb{Z}_2^n$ denote the all-ones vector. Then (\ref{X_and_Z_properties}) implies that \be\label{Z_ket_u} Z^{\otimes n}|u\rangle = (-1)^{d^Tu}|u\rangle\ee for every $u\in\mathbb{Z}_2^n$. Note that $x^2 = x$ for every $x\in\mathbb{Z}_2$. Therefore \be d^Tu = \sum u_i = \sum u_i^2 = u^Tu.\ee If $u\in C$, the self-orthogonality of $C$ implies that $u^Tu = 0$. Hence $d^Tu=0$ for every $u\in C$. Together with (\ref{Z_ket_u}) this implies $Z^{\otimes n}|C\rangle= |C\rangle$. \finpr

\subsection{The expectation value mapping}\label{sect_expectation_value}
Next we prove identity (\ref{partition_expectation_text}). Consider an $n$-qubit system and let $|\pi\rangle=\bigotimes |\pi_i\rangle$ be an $n$-qubit (real or complex) product state, where $|\pi_i\rangle = a_i|0\rangle + b_i|1\rangle$. Define an associated $n$-qubit tensor product operator as follows:
 \be A:= \bigotimes_i A_i, \quad \mbox{where} \quad A_i := \left[ \begin{array}{cc}a_i & b_i \\ b_i & -a_i\end{array} \right].\ee
Now consider an arbitrary self-orthogonal linear code $C$ of length $n$ and the associated $n$-qubit CSS state $|C\rangle$. The following lemma  relates the overlap $\langle C| \pi\rangle$ to the expectation value $\langle C| A|C\rangle$:
\begin{lem}{\bf (Overlap rewriting)}\label{thm_overlap_expectation}
Let $C$ be a binary self-orthogonal linear code of length $n$ and let $|\pi\rangle$ be an $n$-qubit product state. Then \be \sqrt{|C|} \cdot \langle C| \pi\rangle = \langle C| A|C\rangle.\ee
\end{lem}
{\it Proof:} For any $u=(u_1, \dots, u_n)\in\mathbb{Z}_2^n$ we denote \be a^{\bar u} b^{u} = \prod_{i: u_i = 0} a_i \prod_{j: u_j = 1} b_j.\ee Noting that $ZA_i = a_iI + b_i ZX$, we  have \be\label{ZA} Z^{\otimes n}A = \sum_{u\in \mathbb{Z}_2^n} a^{\bar u} b^{u} X(u) Z(u).\ee  We now claim that \be\label{claim} \langle C| X(u)Z(u)|C\rangle = \left \{ \begin{array}{cl} 1 & \mbox{ if } u\in C \\ 0 & \mbox{ otherwise.}\end{array}\right.\ee
To prove the claim, first consider $u\in C$. Then  $u\in C^{\perp}$ since $C$ is self-orthogonal. It follows from (\ref{CSS_stabilizer}) that $X(u)Z(u)|C\rangle = |C\rangle$. Second, consider $u\notin C$. Then for every $v\in C$ we have $u+v\notin C$ as well. It follows that the state \be X(u)Z(u)|C\rangle = \frac{1}{\sqrt{|C|}}\sum_{v\in C} (-1)^{u^Tv}|u+v\rangle\ee is a superposition where each basis state $|u+v\rangle$ corresponds to a bit string lying outside of $C$. This proves the claim.

Since $C$ is self-orthogonal,  we have $Z^{\otimes n}|C\rangle = |C\rangle$ owing to lemma \ref{thm_Z_otimes_n}. With (\ref{ZA}) and (\ref{claim}) it follows that \be \langle C| A|C\rangle &=& \langle C| Z^{\otimes n} A|C\rangle \nonumber \\ &=&  \sum_{u\in \mathbb{Z}_2^n} a^{\bar u} b^{u} \langle C| X(u) Z(u)|C\rangle \nonumber \\ &=& \sum_{u\in C} a^{\bar u} b^{u}.\ee
One the other hand, directly applying the definition of $|C\rangle$ we find \be \langle C|\pi\rangle = \frac{1}{\sqrt{|C|}} \sum_{u\in C} a^{\bar u} b^u. \ee This proves the result. \finpr

We now return to the TCC state $|\Omega\rangle=|S\rangle$. Consider the product state $|\alpha\rangle = \bigotimes |\alpha_a\rangle$ as in (\ref{alpha_a}). The associated tensor product operator  $A= \bigotimes A_a$ is given explicitly by \be\label{A_a} A_a = \frac{1}{ \sqrt{e^{2\beta J_a} + e^{-2\beta J_a}}} \ \left[ \begin{array}{cc}e^{\beta J_a} & e^{-\beta J_a} \\ e^{-\beta J_a} & - e^{\beta J_a}\end{array} \right]\ee
Recalling that $|S|= 2^{F-2}$ (see section \ref{sect_TCC}),  theorem \ref{thm_overlap} and lemma \ref{thm_overlap_expectation} immediately imply:

\begin{thm}{\bf (Expectation value mapping)}\label{thm_Z_expectation}
Consider an arbitrary TCC on a 2-colex. Then \be\label{Z_expectation} {\cal Z}= \frac{\gamma}{\sqrt{2^{F-2}}} \cdot \langle \Omega| A|\Omega\rangle.\ee
\end{thm}

\section{Proof of main result: step 2}\label{sect_proof_main_step2}

In this section we present our classical algorithm to estimate the partition function of the 3-body Ising model. The algorithm is based on the expectation value mapping (\ref{Z_expectation}). The algorithm is described in section \ref{sect_classical_algo}: beforehand,  in section \ref{sect_sampling} we recall some standard theory of Monte Carlo sampling.

\subsection{Chernoff-Hoeffding bound and Sampling}\label{sect_sampling}
The Chernoff-Hoeffding bound is a tool to bound how accurately the expectation value of a random variable may be approximated using sample averages. Let $X_1, \dots X_K$ be i.i.d. real-valued random variables with expectation value $E := \mathbb{E}X_i$ and $|X_i|\leq 1$ for every $i=1, \dots, K$. Then  the Chernoff-Hoeffding bound states that \be\label{hoeff} \mbox{Prob} \left\{ \left|\frac{1}{K}\sum_{i=1}^K X_i - E\right| \leq \epsilon\right\} \geq 1-2 e^{- \frac{K\epsilon^2}{4}}.\ee We will use the Chernoff-Hoeffding bound in the following setting. Let $\{p_x: x\in \mathbb{Z}_2^{\cal V}\}$ be a probability distribution on the set of bit strings. Let $f$ be a real function on $\mathbb{Z}_2^{\cal V}$ with $|f(x)|\leq 1$ and consider the expectation value \be \langle f\rangle = \sum_x p_x f(x).\ee Our goal is to obtain an approximation $c$ of $\langle f\rangle$ by  sampling. To do so, one follows the following standard procedure: first sample $K$ times from $\{p_x\}$, yielding $K$ bit strings $x^1, \cdots, x^K$; then compute the number \be c= \frac{f(x^1)+\cdots + f(x^K)}{K}.\ee The Chernoff bound guarantees that $c$ is $\epsilon$-close to $\langle f\rangle$ with probability at least $p$ provided that the number of samples scales as \be K= O(\frac{1}{\epsilon^2}\cdot \log \frac{1}{1-p}).\ee If the computational cost of a  single sampling run of $\{p_x\}$  is denoted by $N_{\mbox{\scriptsize{samp}}}$ and if the computational cost of computing $f(x)$ on input of $x$ is denoted by $N_{\mbox{\scriptsize{comp}}}$, then the total cost of computing $c$ is \be  [N_{\mbox{\scriptsize{samp}}}+ N_{\mbox{\scriptsize{comp}}}]\cdot O(\frac{1}{\epsilon^2}\cdot \log \frac{1}{1-p}).\ee

\subsection{The algorithm}\label{sect_classical_algo}

Throughout this section we will consider an arbitrary 2-colex and its associated Ising model with partition function ${\cal Z}$. As before, we let ${\cal V}$ and ${ \cal F}$ denote the vertex and face set, and $V$ and $F$ denote the number of vertices and faces of the colex, respectively. The associated Ising model is described in terms of the following data, which are considered to be given as inputs:
\begin{itemize}
\item[(a)] The $V\times F$ vertex-face incidence matrix $B$;
\item[(b)] the inverse temperature $\beta$;
\item[(c)]  the couplings $\{J_a\}$, which may be site-dependent.
\end{itemize}
Additional parameters of the algorithm will be the error $\epsilon$ of the approximation of ${\cal Z}$ and the probability $p$ with which the algorithm succeeds.

Next we present our classical algorithm for estimating ${\cal Z}$. Crucial to our algorithm will be a matrix decomposition which relates the matrices $A_a$, which are real orthogonal matrices, to the (single-qubit) \emph{Clifford group}. Recall that the latter is the group generated by the Hadamard gate $H$ and  the $\frac{\pi}{2}$-gate by $P:=\mbox{ diag}(1, i)$. We will subsequently exploit that the Clifford group acts in a transparent way on TCC states (because the later are stabilizer states) to arrive at our classical algorithm to estimate ${\cal Z}$.

\begin{lem}{\bf (Decomposing orthogonal matrices)}\label{thm_O_2}
Consider a $2\times 2$ real orthogonal matrix $O$. If det$(O)=1$ there exists a (complex) diagonal unitary matrix $D$ such that \be O = P^{\dagger}H D H P.\ee If det$(O)=-1$ there exists a (complex) diagonal unitary matrix $D$ such that \be O = Z P^{\dagger}H D H P.\ee
\end{lem}
{\it Proof:} every $O\in O(2)$ has one of the two following forms, depending on whether its determinant is 1 or $-1$, respectively: \be \left[ \begin{array}{cc}\cos \theta & \sin\theta \\ -\sin\theta& \cos\theta \end{array} \right]\equiv O_+;\quad  \left[ \begin{array}{cc}\cos \theta & \sin\theta \\ \sin\theta& -\cos\theta \end{array} \right]&\equiv& O_-.\nonumber \ee Defining  $D:=\mbox{ diag}(e^{-i\theta}, e^{i\theta})$, the lemma is verified straightforwardly.
\finpr

Here it is interesting to mention that an analogue of the above lemma does not hold for general $U(2)$ matrices i.e. the fact that $O$ is a real orthogonal matrix is crucial.

\begin{thm}{\bf (Classical algorithm for ${\cal Z}$)}\label{thm_classical_algo}
There exists a probabilistic classical algorithm with runtime \be\label{main_runtime2} O( V^3\cdot \frac{1}{\epsilon^2} \cdot \log\frac{1}{1-p})\ee which outputs an estimate ${\cal Z}_{\mbox{\scriptsize{\sc{est}}}}$ of the partition function ${\cal Z}$, such that the inequality \be\label{estimate_classical} |{\cal Z}_{\mbox{\scriptsize{\sc{est}}}}-{\cal Z}|\leq  \frac{\gamma}{ \sqrt{2^{F-2}}} \cdot \epsilon\ee is satisfied with probability at least $p$.
\end{thm}
{\it Proof:} we use lemma \ref{thm_O_2}. Note from (\ref{A_a}) that each $A_a$ is an orthogonal matrix with determinant $-1$. Writing $U:= HP$, lemma \ref{thm_O_2} shows that there exists a diagonal operation \be D_a:=\mbox{ diag}(e^{-i\theta_a}, e^{i\theta_a})\ee  such that $A_a = Z U^{\dagger} D_a U$ for every vertex $a\in {\cal V}$.  Since TCC states are CSS states defined by self-orthogonal classical codes (recall lemma \ref{thm_self_orthogonal}), lemma \ref{thm_Z_otimes_n} implies that $Z^{\otimes V}|\Omega\rangle = |\Omega\rangle$.  Using this last identity together with $A_a = Z U^{\dagger} D_a U$, and denoting \be |\varphi\rangle:= U^{\otimes V}|\Omega\rangle,\ee  it follows that \be\langle \Omega| A |\Omega\rangle &=& \langle \Omega| Z^{\otimes V} A |\Omega\rangle=\langle \varphi| \bigotimes D_a |\varphi\rangle.\label{psi_D_psi}\ee
Denoting
\be f_1(x)&\equiv& \mbox{ Re }\langle x|\bigotimes D_a|x\rangle \nonumber \\f_2(x)&\equiv& \mbox{ Im }\langle x|\bigotimes D_a|x\rangle\nonumber\\ p_x &\equiv& |\langle x|\varphi\rangle|^2,\ee
we have \be\langle \Omega| A |\Omega\rangle &=& \sum p_x f_1(x) + i \sum p_x f_2(x)  \nonumber\\ &=&\langle f_1\rangle + i\langle f_2\rangle.\ee  Since $U$ is a Clifford operation and $|\Omega\rangle$ is a stabilizer state, also $|\varphi\rangle$ is a stabilizer state. It follows from the Gottesman-Knill theorem \cite{Go99} that the distribution $\{p_x\}$ can be sampled in $N_{\mbox{\scriptsize{samp}}}=O(V^3)$ time on a classical computer. Further, its is easily verified that the cost of computing $f_1(x)$ and $f_2(x)$ is $N_{\mbox{\scriptsize{comp}}}= O(V)$. The sampling scheme described in section \ref{sect_sampling} thus allows us to compute numbers $c_1$ and $c_2$ which are $\frac{\epsilon}{2}$-close to $\langle f_1\rangle$ and $\langle f_2\rangle$, respectively, with probability $p$,  with a runtime scaling  \be\label{runtime_classical} O(V^3\cdot \frac{1}{\epsilon^2} \cdot \log\frac{1}{1-p}).\ee  Then $c:= c_1+ic_2$ is $\epsilon$-close to $\langle \Omega| A |\Omega\rangle = \langle f_1\rangle + i\langle f_2\rangle$  with probability $p$. Using the expectation value mapping (\ref{Z_expectation}) it follows that the number \be {\cal Z}_{\mbox{\scriptsize{\sc{est}}}}:= \frac{\gamma }{ \sqrt{2^{F-2}}} c\ee satisfies (\ref{estimate_classical}) with probability $p$, as desired.

\finpr

Note that, interestingly, even though the aim of the algorithm is to estimate a partition function i.e. a sum of positive contributions, the algorithm takes a ``detour'' via the complex numbers; this is done by introducing the diagonal matrices $D_a$.

\section{Quantum simulation}\label{sect_quantum_simulation}

Our quantum simulation algorithm for estimating ${\cal Z}$ will be very simple: it will essentially consist of measuring the expectation value $\langle \Omega|A|\Omega\rangle$ by preparing $|\Omega\rangle$ with a quantum circuit and subsequently measuring the operators $A_a$.

\begin{thm}{\bf (Quantum algorithm for  ${\cal Z}$)}\label{thm_quantum_algo}
There exists a quantum simulation algorithm to estimate ${\cal Z}$ with the same performance (\ref{main_runtime2})-(\ref{estimate_classical}) as the classical algorithm in theorem \ref{thm_classical_algo}.
\end{thm}
{\it Proof:}
Recall the formula (\ref{A_a}) for the matrices $A_a$. It is easily verified that each $A_a$ is a real, symmetric, traceless and orthogonal matrix. Hence its eigenvalues are +1 and $-1$ and there exists a real orthogonal matrix $O_a$ such that $A_a = O_a^T ZO_a$. Computing  the $V$ matrices $O_a$ requires $O(V)$ time resources. Denoting  \be |\xi\rangle := \bigotimes O_a|\Omega\rangle\ee yields \be \langle \Omega| A |\Omega\rangle = \langle\xi| Z^{\otimes V}|\xi\rangle.\ee The quantum algorithm now basically consists of  preparing the state $|\xi\rangle$ and measuring the observable $Z^{\otimes V}$. More precisely, writing \be\label{f_p_quantum} f(x)\equiv \prod_{a\in {\cal V}} (-1)^{x_a}\quad \mbox{and} \quad p_x \equiv |\langle x|\xi\rangle|^2\ee where $x= (x_a: a\in {\cal V})$ denotes a bit string, we have  \be \langle\xi| Z^{\otimes V}|\xi\rangle = \sum_x p_x f(x) =  \langle f\rangle.\ee The cost  $N_{\mbox{\scriptsize{samp}}}$ of sampling the distribution $\{p_x\}$  on a quantum computer is $O(V^3)$: indeed, since $|\Omega\rangle$ is a $V$-qubit stabilizer state (as are all CSS states), there exists an $O(V^3)$ quantum circuit generating this state. Second, applying the operation $\bigotimes O_a$ can be done in $O(V)$ time. Furthermore, the cost $ N_{\mbox{\scriptsize{comp}}}$ of computing $f(x)$ (on input of $x$) is easily shown to be  $O(V)$.  From this point on, the proof is finished by straightforwardly applying the Chernoff bound, analogous to the proof of theorem \ref{thm_classical_algo}. \finpr

\section{Proof of Eqs. (\ref{F_text}) and (\ref{bound_Z_text})}\label{sect_F}

First, the identity  \be F = \frac{V-4g}{2} + 2.\ee follows immediately from (\ref{dimension}).

Second, theorem \ref{thm_Z_expectation} shows that \be\label{geometric} {\cal Z} \leq \frac{\gamma}{\sqrt{2^{F - 2}}}\ee where we have used that  $|\langle \Omega|A|\Omega\rangle|\leq \|A\|\leq 1$ since $A$ is an orthogonal matrix.

\section{High- and low-temperature behavior of ${\cal Z}$}\label{sect_high_low_T}

Here we compute ${\cal Z}$ in two extremal regimes: zero temperature in a ferromagnetic system and  infinite temperature. See the last paragraph in the section \emph{Main results}.

Consider first the zero temperature regime in a ferromagnetic system i.e. $T=0$ and $J_{a}\geq 0$, corresponding to $A_a=Z$ for all vertices $a$.  In this case one has \be \langle \Omega| A|\Omega\rangle= \langle \Omega| Z^{\otimes V}|\Omega\rangle = 1\ee where we have used that $Z^{\otimes V}|\Omega\rangle =  |\Omega\rangle$ owing to lemma \ref{thm_Z_otimes_n}. It follows that ${\cal Z}= \gamma/2^{F/2-1}$.

The high-temperature regime $T=\infty$ corresponds to $A_a = H$ where $H$ is the Hadamard gate. Recalling definition (\ref{S}) of the classical code $S$ and denoting $|+\rangle = \frac{1}{\sqrt{2}}[|0\rangle + |1\rangle]$ one has  \be \langle \Omega| A|\Omega\rangle &=& \langle \Omega| H^{\otimes V}|\Omega\rangle = \sqrt{|S|} \cdot \langle\Omega| +\rangle^{\otimes V} \\ &=& |S|2^{-V/2} = 2^{F -2 - V/2} = 4^{-g}\ee
where: in the second identity we used lemma \ref{thm_overlap_expectation}; in the third identity we used that $\sqrt{|S|}|\Omega\rangle = \sum |s\rangle $ where the sum ranges over all $s\in S$, owing to (\ref{CSS}); in the fourth identity we used $|S|=2^{F-2}$; in the fifth, we used (\ref{dimension}). In conclusion, $\langle \Omega| A|\Omega\rangle = 4^{-g}$ so that ${\cal Z}= \gamma2^{-F/2+1} 4^{-g}$.


\begin{thebibliography}{99}
\bibitem{foot}
Since both color code states and product states can be efficiently prepared by a quantum circuit, the overlap mapping from \cite{Bo08} immediately yields a quantum algorithm to approximate these partition functions (see also below).
\bibitem{foot2}
The quantum algorithms in \cite{Ar10} deal with 2-body Ising models associated with Kitaevs toric code. Their approach is based on tensor network contraction on a quantum computer. It turns out out that this method yields the same result as a quantum algorithm which is directly based on the state overlap mapping \cite{Va07} i.e. the additive errors produced by both approaches are the same, cf. (\ref{deltaolap}). See \cite{Ar10} for details.
\end{thebibliography}
\end{document}